\documentclass[twocolumn,showpacs,epj,amsmath,amssymb,floatfix,eqsecnum]{revtex4-1}
\usepackage{amsmath,amssymb,color}
\usepackage{bm}
\usepackage{epsfig}
\usepackage{psfrag}

\newcommand{\psic}{c}
\newcommand{\bd}{\bm}

\begin{document}

\title{
Non-analytic magnetic field dependence of quasi-particle properties
of two-dimensional metals}

\author{Casper Drukier}
\affiliation{Institut f\"{u}r Theoretische Physik, Universit\"{a}t
  Frankfurt,  Max-von-Laue Str. 1, 60438 Frankfurt, Germany}
\affiliation{Department of Physics, University of Florida, Gainesville, 
Florida 32611, USA}


\author{Philipp Lange}
\affiliation{Institut f\"{u}r Theoretische Physik, Universit\"{a}t
  Frankfurt,  Max-von-Laue Str. 1, 60438 Frankfurt, Germany}
\affiliation{Department of Physics, University of Florida, Gainesville, 
Florida 32611, USA}

\author{Peter Kopietz}
\affiliation{Institut f\"{u}r Theoretische Physik, Universit\"{a}t
  Frankfurt,  Max-von-Laue Str. 1, 60438 Frankfurt, Germany}
\affiliation{Department of Physics, University of Florida, Gainesville, 
Florida 32611, USA}


\date{November 5, 2014}

 \begin{abstract}
We show that
in a weak external magnetic field  $H$ the quasi-particle residue and
the renormalized electron Land\'{e} factor of two-dimensional Fermi liquids
exhibit a non-analytic magnetic field dependence
proportional to $|H|$ which is due to electron-electron interactions and the Zeeman effect.
We explicitly calculate the
corresponding prefactors to second order in the interaction and show
that they are determined by low-energy scattering processes 
involving only momenta close to the Fermi surface.
These non-analytic terms appear in measurable quantities such as the density of states and the
magnetoconductivity.

\end{abstract}

\pacs{71.10.-w, 05.30.Fk, 71.10.Ca}

\maketitle

\section{Introduction}

The non-analytic dependence of thermodynamic susceptibilities 
of  Fermi liquids
on relevant control parameters such as the temperature $T$
or a magnetic field $H$ has recently received 
a lot of attention\cite{Belitz97,Chubukov03,Galitski05,Chubukov05,Betouras05,Chubukov06,Aleiner06,Schwiete06,Shekhter06,Chubukov08,Maslov09,Chesi09,Belitz13}. The non-analytic
corrections are due
to electron-electron interactions and exist for dimensions $D$ in the range
$1 < D \leq 3$. While in three dimensions the leading corrections
to the dominant  analytic terms predicted by the Sommerfeld expansion
are only logarithmic in $T$ and $H$, 
for $D < 3$  the non-analytic corrections to the free energy and its derivatives
scale as  $ T^{D-1}$ and $| H |^{D-1}$.
Recently Belitz and  Kirkpatrick \cite{Belitz13} have shown that
within the framework of the renormalization group
these corrections can be related to the existence of
a certain leading irrelevant coupling constant with scaling dimension $-(D-1)/2$;
a scaling argument then 
gives a simple explanation for the non-analyticities
of thermodynamic quantities as a function of $T$ and $H$
for $1 < D \leq 3$.

The non-analytic magnetic field dependence of the free energy and
the resulting spin susceptibility has been obtained 
by Maslov and Chubukov \cite{Maslov09}. In two dimensions
they found that the magnetic field dependence of 
the grand canonical potential
per unit volume  can be written as
\begin{equation}
 f ( h ) = f (0) - \frac{\chi(0)}{2}  h^2 - \frac{\chi_1}{6}  | h |^3 
 + {\cal{O}} ( h^4 ) .
 \label{eq:free}
 \end{equation}
Here $h = g \mu_B H/2$ is the Zeeman energy of an electron
in a magnetic field $H$, where
$g \approx 2$ is the electron Land\'{e} factor and $\mu_B$ is the Bohr magneton.
To second order in the relevant dimensionless interaction $u = \nu U$
(where $U$ is a short-range bare interaction between two electrons
with antiparallel spin)
the coefficient $\chi_1$ is at zero temperature given by\cite{Maslov09}
 \begin{equation}
 \chi_1 =  2 \nu u^2/E_F + {\cal{O}} ( u^3 ).
 \label{eq:chi1res}
 \end{equation}
Here 
 $\nu = m /(2 \pi)$ is the two-dimensional density of states 
(per spin projection)
at the Fermi energy $E_F$, where $m$ is the electronic mass.
For convenience we use units where $\hbar=k_B =1$ and
normalize the zero-field spin-susceptibility
such that in the non-interacting limit $\chi (0) = 2 \nu$.
The expansion (\ref{eq:free})  of the thermodynamic potential 
implies that the field-dependent spin-susceptibility
 \begin{equation}
 \chi ( h ) = - \partial^2 f ( h ) /\partial h^2  =
 \chi ( 0 )  + \chi_1 | h |  + {\cal{O}} ( h^2 )
 \end{equation}
exhibits a non-analytic magnetic field dependence proportional to~$| h |$.
While in second order perturbation theory
the coefficient $\chi_1$ is positive, to third order in the interaction 
one obtains a negative correction to $\chi_1$,
so that within 
perturbation theory one cannot exclude the possibility
that the sign of $\chi_1$ becomes negative for sufficiently
strong interaction~\cite{Maslov09}.
Although it has been argued \cite{Belitz13}
that this does not happen, 
non-perturbative calculations of $\chi_1$ retaining all
scattering channels are not available.
One can perform a partial resummation to all orders 
by collecting the dominant logarithmic corrections in the Cooper channel, which occur starting
at the third order. In this way the non-analytic dependence 
of the spin-susceptibility on
temperature \cite{Shekhter06}, magnetic field \cite{Maslov09},
or an external  momentum \cite{Chesi09} have been calculated.
In all three cases the coefficient of the non-analytic correction
changes sign if the corresponding parameter is sufficiently small.
The crossover happens at the energy scale associated with the
Kohn-Luttinger instability, but is numerically larger.
It is still likely, though, that the characteristic energy scale
is unmeasurably small.

In this work we show that some quasi-particle properties
such as the quasi-particle residue and the
renormalized Land\'{e} factor also exhibit a non-analytic magnetic field
dependence which in two dimensions is proportional to $ | H |$.
We explicitly calculate the corresponding prefactors 
to leading order in the interaction. We also discuss
consequences for experimentally accessible quantities;
specifically, we show that the tunneling density of states and the
magnetoconductivity both have corrections linear in $| H |$
which arise from the momentum-dependence 
of the self-energy. 

It is important to emphasize that in this work we  will only consider the magnetic field
dependence  arising from the
Zeeman energy of the electronic spin.
We will ignore magnetic field effects associated with the
orbital motion of the electrons, which in an interacting system are also known to generate
non-analytic corrections to various
thermodynamic and transport properties of two-dimensional electrons \cite{Sedrakyan07a,Sedrakyan07b,Sedrakyan08}.
In Sec.~\ref{sec:magnetoconductivity}
we shall further comment on the non-analytic corrections 
arising from the orbital motion in a magnetic field.

\section{Non-analytic magnetic field dependence of 
quasi-particle properties}

Since the non-analytic corrections are due to electron-electron interactions
we ignore the lattice and consider the following Euclidean action
of spin-dependent Grassmann-fields $c_K^{\sigma}$
describing electrons with mass $m$ interacting
with a momentum-independent 
bare interaction $U$ acting between
different spin projections $\sigma_1 \neq \sigma_2$,
 \begin{eqnarray}
  S [ \psic ] &  = &  - \int_{K} \sum_{ \sigma}  [G^{\sigma}_{0} ( K )]^{-1}
  \bar{\psic}_{ K}^{ \sigma} \psic_{K}^{ \sigma} 
 \nonumber
 \\
 & +  & 
 \frac{U}{2}
 \int_{ K} \int_{ K^{\prime}} \int_{Q}   
 \sum_{ \sigma_1 \neq \sigma_2}
 \bar{\psic}_{ K +Q }^{\sigma_1}
 \bar{\psic}_{ K^{\prime}-Q}^{ \sigma_2 }
 {\psic}_{ K^{\prime}}^{ \sigma_2 }
{\psic}_{ K}^{ \sigma_1 }.
 \label{eq:action1}
 \end{eqnarray}
The Gaussian part of the action depends on
the inverse non-interacting Matsubara Green function 
 \begin{equation}
 [G^{\sigma}_{0} ( K )]^{-1} =  i \omega - {\bd{k}^2}/(2m) + \mu +
 \sigma h,
 \end{equation}
where $\mu$ is the chemical potential, $\bd{k}$ is the electronic 
momentum and
$i \omega$ is a fermionic Matsubara frequency.
For convenience we have introduced collective  labels
$K =(  \bd{k} , i \omega )$ and
 $ Q = ( \bd{q} , i \bar{\omega})$, where $i \bar{\omega}$  
is a bosonic Matsubara frequency. We focus on the zero temperature and
infinite volume limit in two dimensions 
where  the integration symbol reduces to
$\int_K = \int  \frac{d^2 k}{(2 \pi)^2 }
 \int \frac{ d \omega}{2 \pi}$.

\begin{figure}[tb]
  \centering
\vspace{7mm}
\includegraphics[width=70mm]{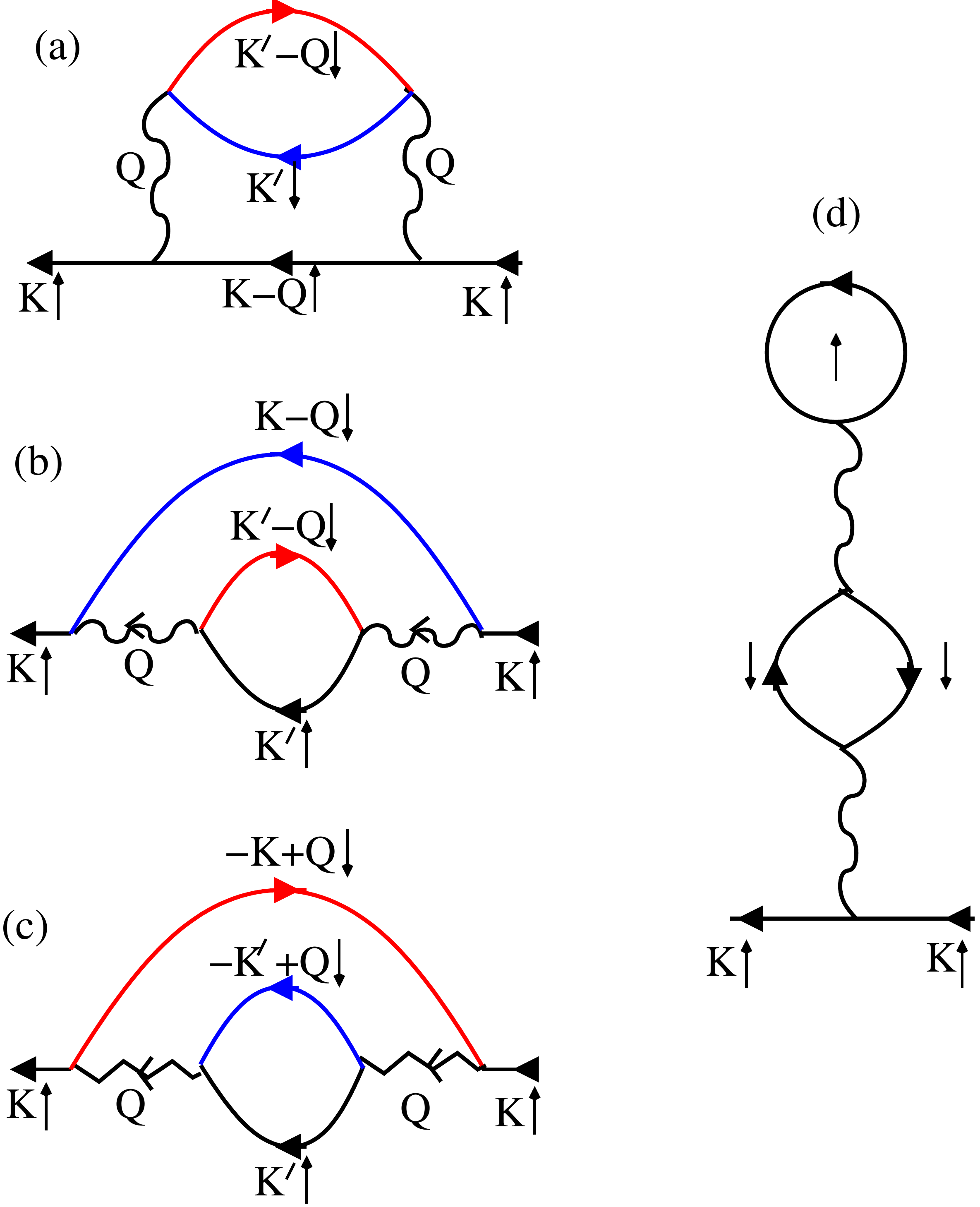}
\vspace{5mm}
  \vspace{-4mm}
  \caption{%
(Color online)
Diagrams (a-c) represent three equivalent ways of
writing the momentum- and frequency-dependent part of the second order 
self-energy $\Sigma^{\uparrow} ( K )$,
corresponding to Eqs.~(\ref{eq:selfa}--\ref{eq:selfc}).
The solid arrows denote the non-interacting  electron Green functions,
while the wavy and zig-zag lines denote the bare interaction.
If the energy-momentum  $Q$ going through the interaction
lines is unrestricted, all three diagrams represent the same
mathematical expression.
However, if we impose a cutoff $\Lambda_0 \ll k_F$ 
on the momentum going  through the interaction lines,
each of these diagrams represents a different low-energy
contribution to the
self-energy, 
corresponding to the three low-energy
processes shown in Fig.~\ref{fig:phasespace}.
For clarity we introduce different symbols for the corresponding
interactions:
(a) forward scattering (wavy line without arrow);
(b) exchange scattering (wavy arrow), and (c) Cooper scattering (zig-zag arrow).
The second order Hartree type of diagram (d) yields a momentum- and frequency independent correction to the self-energy.
}
\label{fig:self2}
\end{figure}

To second order in the interaction,
the momentum- and frequency-dependent part of the 
self-energy of electrons with spin-projection $\sigma$
can be written in the following three equivalent ways
(which can be generated from each other by relabeling and 
changing the order of integrations)
 \begin{subequations}
 \begin{eqnarray}
   \Sigma^{\sigma} ( K )  & = &  - U^2 \int_Q 
 \Pi_0^{- \sigma, - \sigma} (  Q ) G^{\sigma}_0 ( K - Q )
 \label{eq:selfa}
 \\
 & = &
 - U^2 \int_Q 
 \Pi_0^{\uparrow \downarrow} ( \sigma Q ) G^{- \sigma}_0 ( K - Q )
 \label{eq:selfb}
 \\
 & = &      - U^2 \int_Q 
 \Phi_0^{\uparrow \downarrow} (  Q ) G^{-\sigma}_0 ( -K + Q ),
 \label{eq:selfc}
 \end{eqnarray}
\end{subequations}
where the particle-hole and particle-particle bubbles are 
 \begin{subequations} 
\begin{eqnarray}
 \Pi_0^{\sigma  \sigma^{\prime}} ( Q ) & = & \int_K
 G_0^{\sigma} ( K ) G_0^{\sigma^{\prime}} ( K - Q ),
 \\
 \Phi_0^{\sigma  \sigma^{\prime}} ( Q ) & = & \int_K
 G_0^{\sigma} ( K ) G_0^{\sigma^{\prime}} ( -K + Q ).
 \end{eqnarray}
 \end{subequations}
The three different ways of writing the self-energy
in Eqs.~(\ref{eq:selfa}--\ref{eq:selfc}) are 
depicted in Fig.~\ref{fig:self2} (a-c).
In addition, there is a second order self-energy diagram of the Hartree type
shown in Fig.~\ref{fig:self2} (d) which we ignore because
it is independent of momentum and frequency.
A priori it is not clear which of the three expressions
in Eqs.~(\ref{eq:selfa}--\ref{eq:selfc}) is most convenient for
the explicit evaluation of the self-energy. 
However, given the fact that 
the non-analytic magnetic field dependence of the free energy
can be expressed \cite{Maslov09} in terms of the small-momentum part
of the antiparallel-spin particle-hole
bubble $\Pi_0^{\uparrow  \downarrow} ( Q )$,
let us start from the representation (\ref{eq:selfb}).
In two dimensions the relevant
particle-hole bubble can be calculated
exactly at zero temperature. The result can be written as
$\Pi_0^{\uparrow \downarrow } ( Q ) \equiv - \nu 
P ( q, i \bar{\omega} ; h )$ 
with the dimensionless function
    \begin{eqnarray}
 & & P ( q, i \bar{\omega} ; h )  =
1
 - i {\rm sgn} \bar{\omega}  \frac{ k_{F}}{q}  
\Biggl[
 \nonumber
 \\
 & &
 \sqrt{ 1 - \left( \frac{q}{2 k_F }\right)^2 +
 \left( \frac{ \bar{\omega} - 2 i h }{v_F q } \right)^2 - 
 \frac{ i \bar{\omega}}{2 E_F } }
 \nonumber
 \\
  &- &      
 \sqrt{ 1 - \left( \frac{q}{2 k_F }\right)^2 +
 \left( \frac{ \bar{\omega} - 2 i h }{v_F q } \right)^2 + 
 \frac{ i \bar{\omega}}{2 E_F } } \; 
 \Biggr] ,
 \label{eq:bubblenl} 
\end{eqnarray}
where $v_F$ is the Fermi velocity and the Fermi momentum
$k_F = m v_F = \sqrt{ 2 \pi n } $ in the absence of a magnetic field is fixed
by the total density $n$.
Here and below the root symbol denotes the principal branch
of the complex root, defined by ${\rm Re} \sqrt{z} \geq 0$. 
Let us stress  that our result (\ref{eq:bubblenl})
for the antiparallel-spin particle-hole bubble holds for all values of the exchange
momentum $q$; our numerical results using  Eq.~(\ref{eq:bubblenl}) given below
therefore take all types of scattering processes into account, including those 
involving momentum transfers close to $2 k_F$.
Note also that for large $q$ the function $P ( q, i \bar{\omega} ; h )$
vanishes as
 \begin{eqnarray}
 P ( q, i \bar{\omega} ; h )  & \sim & 
 2  (k_F / q)^2,
 \end{eqnarray}
and for large $| \bar{\omega} |$,
 \begin{eqnarray}
 P ( q, i \bar{\omega} ; h )  & \sim & 
  \frac{ 2 h}{ i \bar{\omega} } - \frac{2 h^2}{  \bar{\omega}^2}
 + \frac{ (v_F q )^2}{2 \bar{\omega}^2 } 
 \biggl[ 1 - \Bigl( \frac{q}{ 2 k_F }\Bigr)^2 \biggr] ,
 \hspace{7mm}
 \end{eqnarray}
implying that
the integral (\ref{eq:selfb}) is convergent.

We are interested in the quasi-particle
properties, which are encoded in the  expansion of the
self-energy for small frequencies and for wave-vectors
close to the Fermi surface. Note that for finite $h$ the
Fermi momentum  $k_F^{\uparrow}$ of spin-up electrons
has a different value than the Fermi momentum $k_F^{\downarrow}$
of spin-down electrons.
Given the self-energy $\Sigma^{\sigma} ( k , i \omega )$
(note that the $\bd{k}$-dependence
of the self-energy appears only via $ | \bd{k} | = k$),
the true Fermi surface of electrons with spin-projection $\sigma$
is defined via the solution of
 \begin{equation}
 \frac{ (k_{F}^{\sigma} )^2}{2m} + \Sigma^{\sigma} ( k_F^{\sigma} , 0 ) = \mu + \sigma h.
 \end{equation}
If we neglect the self-energy correction, this yields $(k_F^{\sigma})^2 \approx 
 k_F^2 ( 1 + \sigma h/E_F )$ and to leading order $k_F^{\sigma} \approx k_F + \sigma h / v_F$.
The quasi-particle properties are encoded in the expansion
of the self-energies around the true Fermi surface,
 \begin{eqnarray}
 \Sigma^{\sigma} ( K  ) & \approx & \Sigma^{\sigma} ( k_F^{\sigma} , 0 )
- ( 1 -  Y^{-1}) {\xi}^{\sigma}_{{k}}
 + ( 1 - Z^{-1}) i \omega,
 \hspace{7mm} 
 \label{eq:selfexp}
 \end{eqnarray}
where ${\xi}^{\sigma}_k =  [ k^2 - (k_F^{\sigma})^2 ] /(2m)$.
The fermionic Green function assumes then the quasi-particle form
 \begin{eqnarray}
 G^{\sigma} ( K ) & = & \frac{ 1}{ Z^{-1} i \omega 
 - Y^{-1} \xi^{\sigma}_k }
 \nonumber
 \\
& = & \frac{ Z}{  i \omega - \frac{ k^2 - \bar{k}_F^2}{2 m_{\ast}} + \sigma g_{\ast} h },
 \hspace{7mm}
 \end{eqnarray}
where the effective mass renormalization factor is given by
 \begin{equation}  
 m_{\ast} / m  = Y / Z,
 \end{equation}
while the average Fermi momentum $\bar{k}_F$ and the
renormalization factor $g_{\ast}$ of the electron Land\'{e} factor are
defined by
 \begin{eqnarray}
 \frac{ \bar{k}_F^2}{2m} & = & \mu - \frac{ \Sigma^{\uparrow} ( k_F^{\uparrow} , 0 )
 +  \Sigma^{\downarrow} ( k_F^{\downarrow} , 0 )}{2},
 \\
g_{\ast} & = & \frac{Z}{Y}  \left[ 1 - \frac{ \Sigma^{\uparrow} ( k_F^{\uparrow} , 0 )
 - \Sigma^{\downarrow} ( k_F^{\downarrow} , 0 )}{2h} \right] \notag \\
& \equiv&  \frac{Z}{Y} X.
   \label{eq:Xdef}
 \end{eqnarray}
To calculate the weak-field expansions
of the interaction- and magnetic field dependent  renormalization factors
$X$, $Y$, and $Z$, we  substitute
the exact expression (\ref{eq:bubblenl})
for the particle-hole bubble
into  Eq.~(\ref{eq:selfb})
and perform the angular integration using
 \begin{eqnarray}
 & & \int_0^{2 \pi} \frac{ d \varphi }{2 \pi} \frac{1}{ a + i b + \cos \varphi}
  =   \frac{ {\rm sgn} a}{ \sqrt{ ( a + i b )^2 -1 } }
 \nonumber
 \\
 & = &
\frac{1}{ {\rm sgn}  a \sqrt{ \frac{ R+A}{2}  } + i {\rm sgn } b
\sqrt{ \frac{ R-A}{2}  } },
 \label{eq:integral}
 \end{eqnarray}
where $A = a^2 - b^2 -1$, $B = 2ab$, and $R = \sqrt{A^2 + B^2 }$.
Note that for $ | a | < 1$ the 
 integral (\ref{eq:integral})  has a discontinuity at $b=0$,
so that it is not allowed to commute  the partial derivatives
of the integral with respect to $a$ and $b$ with the angular integration.
In fact, writing ${\rm sgn} b = 2 \Theta (b) -1$ and using the identity \cite{Morris94} 
 \begin{equation}
 \delta ( x ) f ( \Theta ( x ) ) = \delta ( x )  \int_0^1 dt f ( t ),
 \end{equation}
where $ f ( \Theta (x ))$ is an arbitrary function of the
step function $\Theta ( x )$,
it is easy to show that \cite{Bartosch09}
\begin{eqnarray}
& &  \frac{\partial}{\partial (ib ) }
 \int_0^{2 \pi} \frac{ d \varphi }{2 \pi} \frac{1}{ a + i b + \cos \varphi}
\nonumber
 \\
& = & \frac{\partial}{\partial a}
 \int_0^{2 \pi} \frac{ d \varphi }{2 \pi} \frac{1}{ a + i b + \cos \varphi}
-2 \delta (b)  \frac{\Theta ( 1 - a^2 )}{\sqrt{1-a^2}}. \qquad
 \label{eq:discon}
 \end{eqnarray}
After performing the angular integration, the self-energy (\ref{eq:selfb}) 
can be written as
\begin{eqnarray}
 \Sigma^{\sigma} ( k , i \omega ) 
 & = &  - \nu U^2 \int 
\frac{ d {\bar{\omega}}}{2 \pi}  \int_0^{\infty} 
 \frac{ dq q }{2 \pi}  
 P ( q , i \bar{\omega}; \sigma h )
 \nonumber
 \\
 &  & \hspace{-7mm} \times
 \frac{ {\rm sgn} ( \xi_{k} + \frac{q^2}{2m}
 + \sigma h ) }{ \sqrt{
 ( i {\omega} - i \bar{\omega}  - \xi_{k} - \frac{q^2}{2m}
 - \sigma  h)^2 -  ( k q/m )^2  }}, \quad
 \hspace{7mm}
 \label{eq:sigmaint}
 \end{eqnarray}
where $\xi_k = ( k^2 - k_F^2)/(2m)$.
Setting in Eq.~(\ref{eq:sigmaint}) $k = k_F^{\sigma}$ and $i \omega =0$
we obtain $ \Sigma^{\sigma} ( k_F^{\sigma} , 0 )$
and hence the factor $X$  defined Eq.~(\ref{eq:Xdef})
which gives the renormalization of the Land\'{e} factor,
 \begin{eqnarray}
 X & =&  1 +   \frac{\nu U^2}{2h} \int \frac{ d {\bar{\omega}}}{2 \pi} 
 \int_0^{\infty} 
 \frac{ dq q }{2 \pi} \Biggl[ P ( q , i \bar{\omega} ;  h )
 \nonumber
 \\
  & & \times   \frac{   {\rm sgn} ( \frac{q^2}{2m}
 +  2 h ) }{ \bigl[
 ( i \bar{\omega} + \frac{q^2}{2m}
 +   2 h)^2 -  ( v_F^{\uparrow} q )^2  \bigr]^{1/2}} 
 - ( h \rightarrow - h ) \Biggr], \quad
 \label{eq:intX}
 \hspace{7mm}
 \end{eqnarray}
where $v_F^{\sigma} = k_F^{\sigma}/m$.
To calculate the quantities $Y$ and $Z$  defined  via Eq.~(\ref{eq:selfexp}), 
we need the derivatives 
of  the self-energy (\ref{eq:sigmaint}) 
with respect to $\xi_k^{\sigma}$ and  $ i \omega $.
Carefully keeping track of the term arising 
from the discontinuity in Eq.~(\ref{eq:discon}),
we obtain
 \begin{eqnarray}
 \frac{1}{Y}  & = & 1  
 + \nu U^2 \int \frac{ d {\bar{\omega}}}{2 \pi} 
 \int_0^{\infty} 
 \frac{ dq q }{2 \pi} P ( q , i \bar{\omega} ; \sigma h )
 \nonumber
 \\
 & & \times  
 \frac{  ( i \bar{\omega} - \frac{q^2}{2m}
 + 2 \sigma  h  ) {\rm sgn} ( \frac{q^2}{2m}
 + 2 \sigma h ) }{ \bigl[
 ( i \bar{\omega} + \frac{q^2}{2m}
 + 2 \sigma  h)^2 -  ( v^{\sigma}_F q )^2  \bigr]^{3/2}}
 ,
 \label{eq:intY}
 \end{eqnarray}
 \begin{eqnarray}
 \frac{1}{Z}  & = & 1 
 + \nu U^2 \int \frac{ d {\bar{\omega}}}{2 \pi} 
 \int_0^{\infty} 
 \frac{ dq q }{2 \pi} P ( q , i \bar{\omega} ; \sigma h )
 \nonumber
 \\
  &\times & 
 \Biggl[
 \frac{  ( i \bar{\omega} + \frac{q^2}{2m}
 + 2 \sigma  h  ) {\rm sgn} ( \frac{q^2}{2m}
 + 2 \sigma h ) }{ \bigl[
 ( i \bar{\omega} + \frac{q^2}{2m}
 + 2 \sigma  h)^2 -  ( v^{\sigma}_F q )^2  \bigr]^{3/2}}
 \nonumber
 \\
 & &   
 + 2 \delta ( \bar{\omega} ) 
 \frac{ \Theta ( (v_F^{\sigma}  q )^2 - ( \frac{q^2}{2m} + 2 \sigma h )^2 )}{
 \bigl[  (v_F^{\sigma} q )^2 - ( \frac{q^2}{2m} + 2 \sigma h )^2 \bigr]^{1/2}}
 \Biggr] .
 \label{eq:intZ}
 \hspace{7mm}
 \end{eqnarray}
The remaining two
integrations over $q$ and $\bar{\omega}$ can be done numerically.
By  plotting the results versus $h$ and fitting the $h$-dependence to straight lines
we find that for small $h$ all three coefficients $X$, $Y$, and $Z$
have a non-analytic term linear in $|h |$,
 \begin{subequations}
 \begin{eqnarray}
 X (h ) & = & X (0) + X_1 | h |  + {\cal{O}} ( h^2 ),
 \label{eq:gh}
 \\
  Y ( h ) & = & Y (0) + Y_1 | h | + {\cal{O}} ( h^2 ),
 \label{eq:Yh}
 \\
 Z ( h ) & = & Z (0) + Z_1 | h | + {\cal{O}} ( h^2 ), 
 \label{eq:Zh}
 \end{eqnarray}
 \end{subequations}
where, with a numerical accuracy of the order of one percent,
all prefactors of the non-analytic terms have the same numerical value,
 \begin{equation}
 X_1 \approx Y_1 \approx Z_1 \approx 0.50 \frac{u^2}{E_F}
 + {\cal{O}} ( u^3 ).
 \label{eq:XYZres}
 \end{equation}
Here $u = \nu U$ is the relevant dimensionless interaction.
Moreover, from our numerical evaluation of the integrals 
in Eqs.~(\ref{eq:intX}-\ref{eq:intZ}) we find that
the numerical values of $Y_1$ and $Z_1$
are determined by exchange-momenta $q$ in the entire interval  $q \in [0, 2 k_F]$, 
indicating that other scattering processes than exchange scattering contribute.
On the other hand, 
the value of $X_1$ is completely determined by the small-$q$ part
of $\Pi_0^{\uparrow \downarrow } ( Q )$,
which is also the case for
the susceptibility coefficient $\chi_1$ in Eq.~(\ref{eq:chi1res}).

\begin{figure}[tb]
  \centering
\includegraphics[width=47mm]{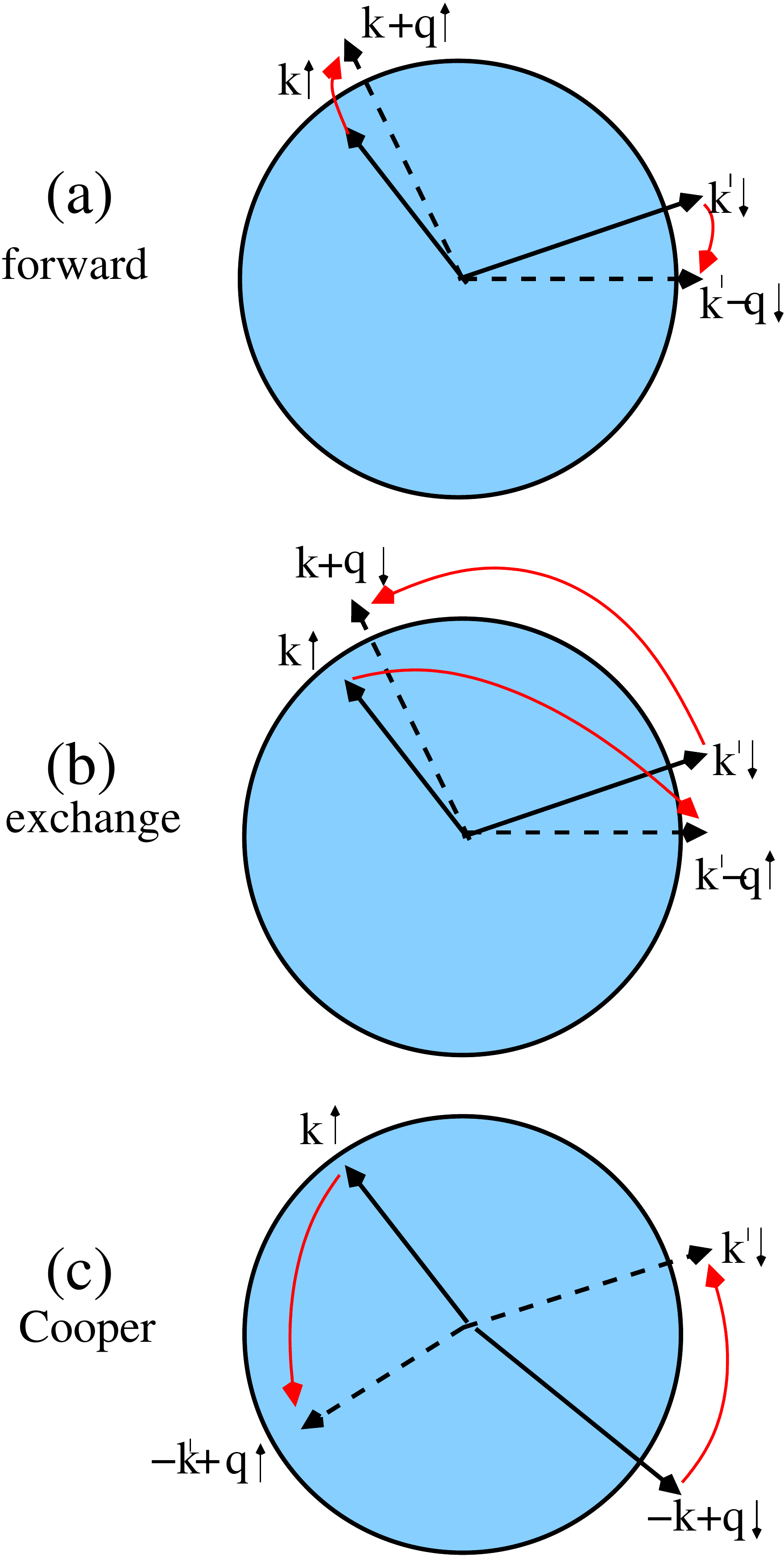}
\vspace{5mm}
  \vspace{-4mm}
  \caption{%
(Color online)
Graphical representation of the 
three different types of scattering processes 
in our effective low-energy interaction (\ref{eq:Sphasespace}) in two dimensions.
All momenta are assumed to be close to the Fermi surface
(represented by a shaded circle); the momentum $|\bd{q}|$
is assumed to be small compared with $k_F$.
Solid (dashed) arrows denote the momenta before (after) 
the scattering process, while curved arrows 
connecting states with the same spin projection denote the 
momentum transfer associated with the scattering processes.
}
\label{fig:phasespace}
\end{figure}

The identity $Y_1 = Z_1 $
implies that, at least to this
order in the interaction, the effective mass $m_{\ast}/ m = Y / Z$ 
does not exhibit 
a linear dependence on the magnetic field.
In order to prove this
and to classify the 
various contributions to the non-analytic terms, let us anticipate that
the numerical values of the coefficients $X_1$, $Y_1$, and $Z_1$ are completely
determined by low-energy scattering processes involving momenta
in the vicinity of the Fermi surface.
This implies that these coefficients
can be calculated from an effective low-energy model
containing only states with momenta in a thin shell around the Fermi surface.
It turns out that in two dimensions only the three types
of low-energy scattering processes shown in Fig.~\ref{fig:phasespace}
are possible~\cite{Belitz97,Shankar94}.
Let us therefore replace 
the bare interaction in our original model~(\ref{eq:action1}) by an effective
low-energy interaction containing only 
scattering processes with momenta in the vicinity of the Fermi surface.
In two dimensions the corresponding Euclidean action 
can be written as \cite{Belitz97}
 \begin{eqnarray}
 S_{\rm int} [\psic ] & \approx & \frac{1}{2} \int_Q 
 \Theta ( \Lambda_0 - | \bd{q} | ) \sum_{\sigma_1 \neq \sigma_2}
 \Bigl[  U_f \bar{D}_{ Q}^{ \sigma_1 \sigma_1 } D_{ Q }^{ \sigma_2 \sigma_2} 
 \nonumber
 \\
 &  & - U_x \bar{D}_{ Q}^{ \sigma_1 \sigma_2 } D_{Q }^{ \sigma_1 \sigma_2}
  +  U_{c}  \bar{C}_{ Q}^{\sigma_1 \sigma_2 } C_{Q }^{ \sigma_1 \sigma_2}
 \Bigr],
 \label{eq:Sphasespace}
 \end{eqnarray}
where $ \Lambda_0 \ll k_F$ is an ultraviolet cutoff and
we have introduced the bilinear composite fermion fields
 $
 D_Q^{\sigma_1 \sigma_2}  =  
 \int_K \bar{\psic}_{ K}^{ \sigma_1}  \psic_{ K + Q}^{ \sigma_2}$ and
$ C_Q^{\sigma_1 \sigma_2} =
 \int_K {\psic}_{ - K  }^{ \sigma_1}  \psic_{ K + Q}^{ 
 \sigma_2} $.
Our original interaction in Eq.~(\ref{eq:action1}) corresponds to
$U_f = U_x = U_c = U$, but it is instructive
to introduce separate interaction constants
$U_f$ (forward scattering), $U_x$ (exchange scattering), and $U_c$
(Cooper scattering). 
For our low-energy model the three contributions to the
self-energy  shown in Fig.~\ref{fig:self2} are additive, so that to second order
in the interaction we may approximate
 \begin{equation}
  \Sigma^{\sigma} ( K ) \approx \Sigma^{\sigma}_f ( K ) + 
 \Sigma^{\sigma}_x ( K ) +  \Sigma^{\sigma}_c ( K ) ,
 \end{equation}
with
 \begin{subequations}
 \begin{eqnarray}
   \Sigma^{\sigma}_f ( K )  & = &  - U_f^2 \int_Q \Theta ( \Lambda_0 - | \bd{q} | )
 \Pi_0^{- \sigma, - \sigma} (  Q ) G^{\sigma}_0 ( K - Q ),
 \nonumber
 \\
 & &
 \label{eq:selfaa}
 \\
  \Sigma^{\sigma}_x ( K ) 
 & = &
 - U_x^2 \int_Q  \Theta ( \Lambda_0 - | \bd{q} | )
 \Pi_0^{\uparrow \downarrow} ( \sigma Q ) G^{- \sigma}_0 ( K - Q ),
 \nonumber
 \\
 & &
 \label{eq:selfbb}
 \\
 \Sigma^{\sigma}_c ( K ) 
 & = &      - U_c^2 \int_Q  \Theta ( \Lambda_0 - | \bd{q} | )
 \Phi_0^{\uparrow \downarrow} (  Q ) G^{-\sigma}_0 ( -K + Q ).
 \nonumber
 \\
 & &
 \label{eq:selfcc}
 \end{eqnarray}
\end{subequations}
Due to the
ultraviolet cutoff $\Lambda_0 \ll k_F$ imposed on the momentum carried
by the interaction lines there is no double counting.
The replacement of the original interaction (\ref{eq:action1})
by the low-energy interaction (\ref{eq:Sphasespace})
is analogous to an established procedure in one dimension, where
the low-energy physics of lattice models
can be described by the so-called g-ology model containing
only scattering processes involving momenta close to the
Fermi points \cite{Solyom79}.
Although the g-ology model depends implicitly on an
ultraviolet cutoff $\Lambda_0 \ll k_F$ defining the
width of the relevant shell around the Fermi points,
universal low-energy properties are independent of
$\Lambda_0$. The model (\ref{eq:Sphasespace}) can be considered as a 
two-dimensional
analogue of the  $g$-ology model.
If we  calculate the
field-independent parts $X (0)$, $Y ( 0 )$, and $Z (0)$
of our renormalization factors
within our low-energy model, we find
that these quantities explicitly depend on
the ultraviolet cutoff $\Lambda_0$ inherent in the definition of the model.
On the other hand, the coefficients 
$X_1$, $Y_1$, and $Z_1$ of the terms linear in $|h |$ are
cutoff-independent.

Because in our low-energy model 
$| \bd{q} | \leq \Lambda_0 \ll k_F$, 
we may approximate the
bubbles in the self-energy diagrams by their limits for  small momenta $q \ll k_F$
and for frequencies $| \bar{\omega} | \ll E_F$.
In this regime, the antiparallel-spin
particle-hole bubble given in
Eq.~(\ref{eq:bubblenl}) simplifies~\cite{Maslov09},
 \begin{equation}
 \Pi_0^{\uparrow \downarrow } ( Q ) 
 \approx  - \nu \left[
   1 - \frac{ | \bar{\omega} | }{
 \sqrt{ ( \bar{\omega} - 2 i h )^2 +  (v_F q )^2 }} \right],
 \label{eq:Pilin}
 \end{equation}
while the parallel-spin particle-hole bubble
$\Pi_0^{\sigma \sigma} ( Q ) $ can be obtained by setting $h=0$
on the right-hand side of Eq.~(\ref{eq:Pilin}).
We neglect the magnetic field dependence of the Fermi velocity because
it can only give rise to corrections which are analytic in $h$.
Within the same approximation, the small-$q$ limit of the particle-particle
bubble is 
 \begin{eqnarray}
& & \Phi^{\uparrow \downarrow}_0 ( Q )   
\approx
 \nu \ln 
 \left| \frac{  2 \Omega_0}{  
\bar{\omega}  - 2 i h +  {\rm sgn} \bar{\omega} 
 \sqrt{ ( \bar{\omega} - 2 i h )^2 + (v_F q )^2 } } \right|,
 \nonumber
 \\
 & &
 \label{eq:phipp}
 \end{eqnarray}
where $\Omega_0 \lesssim E_F$ is an ultraviolet cutoff which restricts the integration to
intermediate states in a thin shell $ | \xi_{k} |  \leq \Omega_0 $ 
around the Fermi surface.
The integrations appearing in the 
expressions for $Y_1$ and $Z_1$ 
are ultraviolet convergent so that we may take the limit  $\Lambda_0 
 \rightarrow \infty$ and $\Omega_0 \rightarrow \infty$. 
Then the relevant  integrals  can be performed analytically and we finally
obtain for the universal coefficients of the terms linear in $| h |$,
 \begin{eqnarray}
 Y_1 &=&  Z_1  =     \frac{u_x^2  +   u_c^2 }{4 E_F}   ,
 \label{eq:Yallres}
 \end{eqnarray}
where $u_i = \nu U_i$. Note that forward scattering processes do 
not contribute to the above non-analytic corrections. 
Setting $u_x = u_c =u$ in Eq.~(\ref{eq:Yallres}), we 
obtain agreement with our direct numerical evaluation
in Eq.~(\ref{eq:XYZres}), so that for this type of interaction
the identity $Y_1 = Z_1$ is established exactly.

The evaluation of the coefficient $X_1$
is more tricky, because the 
integral (\ref{eq:intX}) decays more slowly than the integrals
defining $Y_1$ and $Z_1$. In practice, we first
calculate $ \partial \Sigma^{\sigma} ( k_F^{\sigma} , 0 ) / \partial h $
and integrate the resulting expression over $h$ 
to recover the difference 
in Eq.~(\ref{eq:intX}). Unfortunately, our low-energy
approximation (\ref{eq:phipp}) for the particle-particle bubble
with magnetic-field independent Fermi surface cutoff $\Omega_0$
seems  not to be sufficient to extract the complete non-analytic $h$-dependence
of the coefficient $X ( h )$.
However, we know from our direct numerical
evaluation of Eq.~(\ref{eq:intX}) that
the relevant integral is completely
determined by small exchange momenta $q \ll k_F$, as discussed
in the text
after Eq.~(\ref{eq:XYZres}).
To evaluate $X_1$ within our low-energy model,
it is therefore sufficient to take only the
contribution from the exchange scattering channel into account.
After a straightforward calculation we obtain
 \begin{eqnarray}
X_1 & = &    \frac{  u_x^2   }{2 E_F},  
 \end{eqnarray}
which agrees with Eq.~(\ref{eq:XYZres}) if we set $u_x = u$.

%
%

We conclude that
the coefficients $X_1$, $Y_1$ and $Z_1$ of the terms linear in $| h |$ are indeed
determined by low-energy processes involving momenta close to the
Fermi surface, as anticipated.
Using the methods developed in Ref.~[\onlinecite{Maslov09}],
we can show analytically that the numerical values of these coefficients
are determined by scattering processes involving momentum
transfers in the entire range  $[ 0 , 2 k_F]$.
As a consequence, there are no correlations between the
momenta $\bd{k}$ and $\bd{k}^{\prime}$
of the two incoming electrons
shown in Fig.~\ref{fig:phasespace}, 
so that for a momentum-dependent interaction 
$X_1$, $Y_1$, and $Z_1$
will depend on all angular harmonics of the interaction on the Fermi surface.
This result should be contrasted with the well-known\cite{Fujimoto90,Halboth98,Chubukov03,Chubukov05} non-analytic behavior of the real part of the self-energy
on the real frequency axis  on resonance ($\omega = \xi_{\bd{k}}$)
for vanishing magnetic field, which is proportional to
$ \omega | \omega |$,
with a prefactor  that depends exclusively on scattering processes 
involving momentum transfers $q =0$ and $q =2 k_F$.
Hence, for real frequencies and finite magnetic field
the non-analytic dependence of $\Sigma ( \bd{k} , \omega )$ 
on $\omega $ and $h$ must be described by a non-trivial
cross-over function depending on the ratio $| \omega | / | h |$ which takes the different
nature of the dominant scattering processes
in the two limits  $| \omega | / | h | \ll 1$ and
 $| \omega | / | h | \gg 1$  into account.

\section{Non-analytic corrections to measurable quantities}
\label{sec:exp}

We now discuss experimental observables which are sensitive to the
non-analytic magnetic field dependence of the quasi-particle
properties. The quasi-particle residue and
the renormalized single-particle dispersion can be determined
via  photoemission experiments, which would be the most direct method
to reveal the non-analytic magnetic field dependence of $Z$
 predicted in this work.
Due to the identity $Y_1 = Z_1$, the effective mass
$m_{\ast}/ m = Y/Z$  does not exhibit any non-analytic magnetic field dependence to second
order in the interaction. Whether this remains true also to higher order in the interaction
remains an open problem.

\subsection{Tunneling density of states}

The non-analytic magnetic field dependence of the 
renormalization factor $Y ( h ) $ associated with the
momentum-derivative of the self-energy determines
the renormalized density of states $\nu_{\ast} (h ) $, which
in the quasi-particle approximation is given by
 \begin{equation}
 \nu_\ast (h ) =  Z \int \frac{d^2 k}{ (2 \pi )^2 }  \delta \left( 
  \frac{ k^2 - \bar{k}_F^2}{2m_{\ast}} -
  g_{\ast} h \right) = Y (h )  \frac{m}{2 \pi}.
 \end{equation} 
Hence, the magnetic field dependence of the
renormalized density of states is given by
 \begin{eqnarray}
 \frac{\nu_{\ast} (h ) - \nu_{\ast} (0)}{\nu} & = & Y_1 | h| 
 + {\cal{O}} ( h^2)
 \nonumber
 \\
 & = & \frac{ (u_x^2 +  u_c^2 ) |h |}{4
 E_F} + {\cal{O}} ( h^2, u^3 | h | ).
 \hspace{7mm}
 \label{eq:nures}
 \end{eqnarray}
Note that the renormalized density of states can 
be measured via tunneling experiments. In principle, it should be possible
to verify Eq.~(\ref{eq:nures}) for weakly interacting two-dimensional Fermi liquids with very low disorder;
we are not sure, however, if the required accurary can be achieved with
currently available equipment.

\subsection{Magnetoconductivity}
 \label{sec:magnetoconductivity}

\begin{figure}[tb]
  \centering
\vspace{7mm}
\includegraphics[width=80mm]{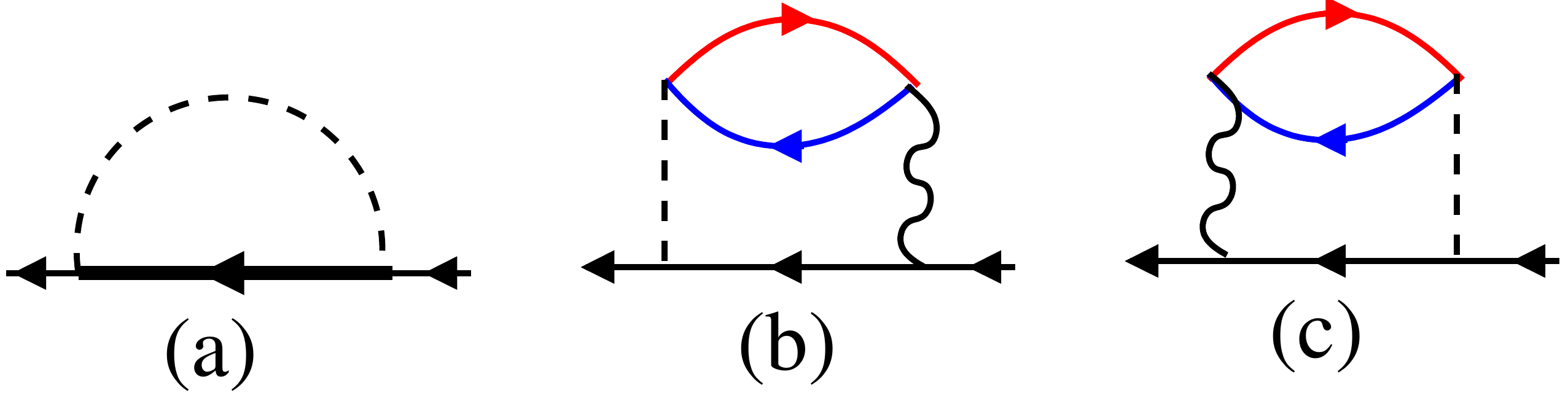}
\vspace{5mm}
  \vspace{-4mm}
  \caption{%
(Color online)
The diagram (a) represents the self-energy due to impurity scattering
in self-consistent Born approximation with quasi-particle propagator, 
see Eq.~(\ref{eq:selfimp}).
The dashed line represents the impurity correlator  while the thick solid
arrow represents the disorder averaged quasi-particle propagator given in
Eq.~(\ref{eq:Gimp}).
The diagrams (b) and (c) represent first order (in the interaction) 
corrections to the self-energy due to impurity scattering.
The imaginary part of these  diagrams generates a non-analytic correction
to the renormalized impurity scattering time $\tau_{\ast}$.
In Appendix  A we show that these diagrams
contribute to the
weak coupling behavior of the magnetoconductivity 
in the ZNA-formula (\ref{eq:ZNA}) to first order in the interaction.
}
\label{fig:impurity}
\end{figure}

\begin{figure}[tb]
  \centering
\vspace{7mm}
\includegraphics[width=80mm]{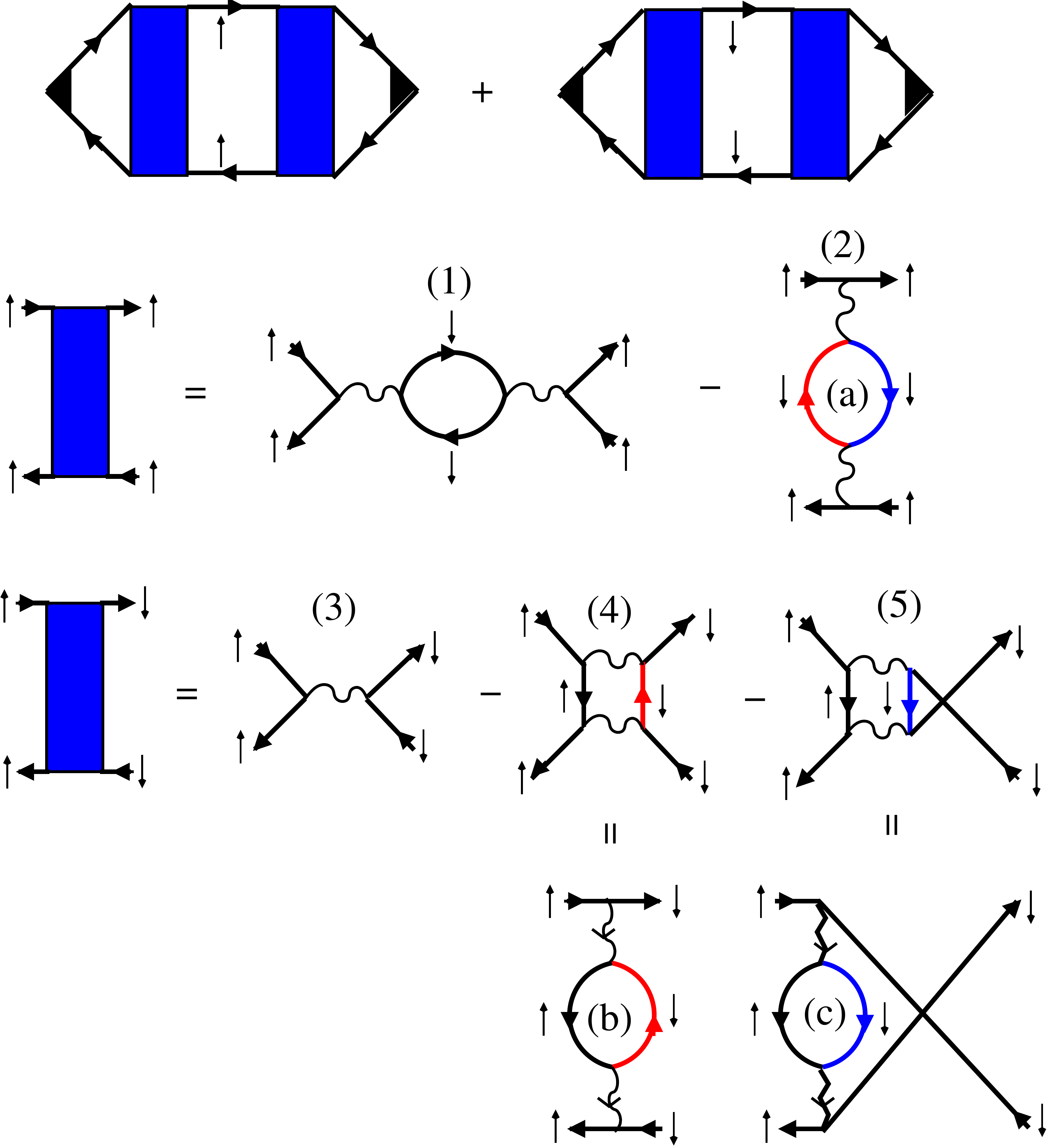}
\vspace{5mm}
  \vspace{-4mm}
  \caption{%
(Color online)
First line: diagrammatic representation of the Kubo formula (\ref{eq:Kubo})
for the conductivity of a Fermi liquid.
The solid arrows denote the disorder averaged quasi-particle  Green functions defined
in Eq.~(\ref{eq:Gimp}), the black triangles denote the bare current vertices,
and the shaded boxes represent  vertex corrections.
In the second and third line we give the perturbative expansions of the
vertex corrections up to second order in the interaction.
When inserted into the conductivity bubble, the diagrams (4) and (5)
generate the so-called Aslamazov-Larkin corrections to the conductivity.
In the last line we redraw these diagrams using the notation introduced
in Fig.~\ref{fig:self2}.
If we impose an ultraviolet  cutoff $\Lambda_0 \ll k_F$ 
on the momentum carried by the interaction lines,
the diagrams (a-c)  generate  in the Kubo formula (\ref{eq:Kubo})
the self-energy corrections shown in Fig.~\ref{fig:self2} (a-c);
we use the same color coding to identify the corresponding bubbles.
The vertex corrections described by the 
second order diagram (1) and the first order diagram (3) correspond 
to  momentum- and frequency-independent
self-energies which do not contribute
to the conductivity (\ref{eq:Kubo}).
}
\label{fig:kubo}
\end{figure}

It turns out that  the factor
 $Y (h)$ appears also in the 
 magnetoconductivity $\sigma (h)$ via the
renormalization of the current vertices in the Kubo formula.
In order to obtain a finite conductivity, we add elastic impurity
scattering to our model.
In the quasi-particle approximation the
disorder averaged fermionic Green function is then
 \begin{equation}
 G^{\sigma} ( K ) = \frac{ Z}{ i \omega 
 - ( m/ m_{\ast} ) \xi_k^{\sigma}   + i {\rm sgn \omega }/(2 \tau_\ast ) } .
 \label{eq:Gimp}
 \end{equation}
To determine the inverse scattering time $1/ \tau_{\ast}$,
we calculate the contribution   $\Sigma_{\rm imp} ( i \omega )$ to the self-energy
due to impurity scattering  for short-range disorder within the Born approximation
using the quasi-particle approximation for the propagator
in the loop integral. Evaluating the corresponding Feynman diagram shown
in Fig.~\ref{fig:impurity} (a) we obtain
  \begin{equation}
 \Sigma_{\rm imp} ( i \omega ) =  - i    {\rm sgn} \omega  \frac{Y  }{2 \tau }  ,
 \label{eq:selfimp}
  \end{equation}
where $\tau$ is the elastic lifetime  in the absence of interactions.
In this approximation, the renormalized inverse scattering time is given by 
\begin{equation}
 1/ \tau_{\ast} =  - 2 Z \,  {\rm Im}   \Sigma_{\rm imp} ( i  \eta)  = ZY / \tau,
 \end{equation}
where $\eta >0$ is infinitesimal.
Using the Drude formula for the  conductivity of a Fermi liquid
with effective mass $m_{\ast}$ and scattering time $\tau_{\ast}$ we 
obtain \cite{Pines66}
\begin{eqnarray}
 \sigma ( h )  & = &  \frac{ n e^2 \tau_{\ast}}{m_{\ast}} =
\frac{\sigma_0}{Y^2} .
 \label{eq:conres}
 \end{eqnarray}
Here $\sigma_0 = n e^2 \tau /m$ is the conductivity without interactions 
and $n = k_F^2 /( 2 \pi )$ is the total electronic density.
Alternatively, Eq.~(\ref{eq:conres}) can be obtained from the Kubo formula
for the conductivity of a $D$-dimensional Fermi  liquid
 \begin{eqnarray}
 \sigma & = & \frac{e^2 }{2 \pi } \frac{ v_F^2}{D}
 \sum_{\sigma}   \left[ 1 + \left. \frac{ \partial \Sigma^{\sigma} ( k , 0 )}{\partial \xi_k^{\sigma} }
  \right|_{ \xi_k^{\sigma} =0} \right]^2 
 \nonumber
 \\
 & & \times 
 \int \frac{ d^D k}{ (2 \pi )^D} G^{\sigma} ( \bd{k} , i \eta ) G^{\sigma} ( \bd{k} , -i  \eta ) .
 \label{eq:Kubo}
 \end{eqnarray}
For vanishing magnetic field,
this formula has first been derived  by Langer \cite{Langer61} 
and has been justified by several other
authors \cite{Eliashberg61,Michaeli09,Maslov12}.
A diagrammatic representation of Eq.~(\ref{eq:Kubo}) is shown
in the first line of Fig.~\ref{fig:kubo}. 
The  factor containing the momentum-derivative of the self-energy 
in the Langer-formula~(\ref{eq:Kubo})  takes 
the low-energy part of vertex corrections due to interactions
into account,  as explained in the caption of Fig.~\ref{fig:kubo}.
The fact that vertex corrections in the Kubo formula can
be expressed in terms of the momentum-derivative of the self-energy
is guaranteed by a Ward identity~\cite{Langer61,Yamada04}.
The vertex corrections corresponding to the three low-energy expressions
for the second order self-energy given in Eqs.~(\ref{eq:selfaa}-\ref{eq:selfcc}) are given by the diagrams labeled (a-c) in Fig.~\ref{fig:kubo}.
Combining Eqs.~(\ref{eq:Yh}, \ref{eq:Yallres}) and (\ref{eq:conres})
we obtain, at second order in the interaction and for weak magnetic fields,
a negative magnetoconductivity (positive magnetoresistance)
proportional to $ | h |$ which is 
given by
 \begin{equation}
 \frac{\sigma (h )_{  \rm{L}2 } - 
\sigma (0)_{ \rm{L}2 }}{\sigma_0} = - \frac{( u_x^2  +  u_c^2 )| h |}{2 E_F}+
{\cal{O}} ( h^2, u^3). 
 \label{eq:condme}
 \end{equation}
where the subscript ${\rm{L}2}$ indicates 
that this expression has been obtained
from the Langer formula (\ref{eq:Kubo}) 
for the conductivity using second order perturbation theory
for the renormalized current vertex.

It is instructive to compare this result
with the expression for the magnetoconductivity
derived by Zala, Narozhny and Aleiner~\cite{Zala01} (ZNA)
for a two-dimensional metal in the ballistic high-field regime
$1/\tau \ll T \ll h \ll E_F$. 
In our notation the ZNA-result
for the magnetoconductivity
can in this regime be written as
\begin{equation}
 \frac{\sigma (h )_{\rm ZNA} - \sigma (0)_{\rm ZNA}}{\sigma_0} =  \frac{| h |}{ E_F}
 \frac{ 2 F_0^{\sigma}}{1 + F_0^{\sigma}} g ( F_0^{\sigma} ),
 \label{eq:ZNA}
 \end{equation}
where the dimensionless Landau parameter
 $F_0^{\sigma}$ specifies the strength of the spin-exchange interaction,
and the function 
$ g ( F_0^{\sigma} )$ has for small $F_0^{\sigma}$
the expansion\cite{Zala01}
 \begin{equation}
  g ( F_0^{\sigma} ) = 1 - 
 \left(  \frac{5}{4} - \ln 2 \right) F_0^{\sigma} +
 {\cal{O}} ( (F_0^{\sigma} )^2 ).
 \end{equation}
 
The ZNA-formula (\ref{eq:ZNA}) has been obtained from a proper disorder
average of the Kubo formula,
using  the random-phase approximation for the effective interaction in the
exchange scattering channel.
High-energy processes involving momenta which are not close to the
Fermi surface are taken into account phenomenologically
via the Landau parameter $F_0^{\sigma}$.
Obviously, Eqs.~(\ref{eq:condme}) and (\ref{eq:ZNA}) both
predict that the leading magnetic field dependence
of the magnetoconductivity is linear in $| h |$.
However, at weak coupling where the Landau parameter $F_0^{\sigma}$
is to leading order proportional to the negative bare interaction $-u_x$ 
in the  exchange scattering channel, 
the ZNA-result is linear in $u_x$, whereas our Eq.~(\ref{eq:condme})  
is quadratic in the low-energy couplings $u_x$ and $u_c$.
This discrepancy is simply
due to the fact that in our work we have not considered
first order interaction processes of the type shown in
Fig.~\ref{fig:impurity} (b) and (c). In Appendix A we show explicitly 
how to recover the leading weak coupling behavior of the ZNA-formula (\ref{eq:ZNA})
from the disorder averaged Kubo formula; we also show
that the Langer-formula  (\ref{eq:Kubo}) does not take the leading
vertex corrections due to impurity scattering in the Kubo formula into account.

Although the weak-coupling expansion of the ZNA-formula (\ref{eq:ZNA}) generates a 
term of order  $(F_0^{\sigma})^2 = u_x^2$,
it is not obvious whether this term includes
the non-analytic corrections to the current vertices
which can be related to the momentum-derivative of the self-energy.
In any case, the ZNA-formula (\ref{eq:ZNA}) is incomplete
because ZNA have ignored the Cooper channel, which according
to Eq.~(\ref{eq:condme}) contributes to the weak-coupling expansion of
the magnetoconductivity
on equal footing with the exchange channel.
On the other hand, our use 
of the Langer-formula (\ref{eq:Kubo}), 
which is by itself on solid 
grounds \cite{Langer61,Eliashberg61,Michaeli09, Maslov12}, 
in a situation where quantum-interference effects \cite{Zala01}  produce
non-analytic terms in $H$ and $T$  is not completely justified, as discussed in 
Appendix A. 
It is therefore likely that other  second order 
processes
in the perturbative expansion of the magnetoconductivity, 
which are not properly taken into account via the Langer formula,
will give rise to additional contributions of order $u^2 | h |$ 
to the magnetoconductivity.

 
Combining our result (\ref{eq:condme}) for the magnetoconductivity 
with the weak-coupling expansion of the ZNA-formula,
we conclude that
the non-analytic magnetic-field dependence of the magnetoconductivity for small magnetic fields
and for small values of the short-range bare interaction $u$ is 
of the form 
\begin{eqnarray}
\frac{\sigma(h)-\sigma(0)}{\sigma_0} & =&\frac{|h|}{E_F} \left[ -2 u
+ c_{2} u^2 + {\cal{O}} ( u^3 ) \right]
 \nonumber
 \\
 & &  
+ {\mathcal{O}(h^2, |h|^{1/3} u^3  )}.
\label{eq:condfullres}
\end{eqnarray}
The calculation of the
numerical value of the second order coefficient $c_2$
is beyond the scope of this work. Note that $c_2$ cannot be
obtained from the weak coupling expansion of the ZNA-formula (\ref{eq:ZNA})
because this formula neglects the contribution from the Cooper channel
which according to Eq.~(\ref{eq:condme}) does contribute to $c_2$.
Unfortunately $c_2$ can also not be extracted from
our Eq.~(\ref{eq:condme}), because our simple 
calculation using the Langer formula
does not amount to a systematic
expansion of the disorder averaged Kubo formula to second order in
the interaction.
 Let us also note that  according to Sedrakyan and 
co-authors \cite{Sedrakyan07a,Sedrakyan07b,Sedrakyan08}, 
at the third order in the interaction, orbital magnetic-field effects yield
a more singular $|h|^{1/3}$-type non-analyticity,
which is responsible for the correction of order $|h|^{1/3} u^3$
in Eq.~(\ref{eq:condfullres}).
However, for weak interactions  this correction
is only dominant for very small magnetic fields, $ | h |/ E_F \ll u^3 \ll 1$.
In the regime $u^{3/2} \ll | h | / E_F \ll 1$ the
term  $c_2 u^2$ in  Eq.~(\ref{eq:condfullres}) is the dominant
correction to the leading linear in $u$ behavior of the magnetoconductivity. The condition
$u^{3/2}\ll |h|/E_F \ll 1$ can possibly be satisfied in the 2D electron gas realized in doped
semiconductor heterostructures based on GaAs. For typical densities the Fermi energy in these
systems is of the order of $10\,\textrm{meV}$ (see  Refs.~[\onlinecite{Tsukernik01,Eshkol06}]), 
so that $|h|/E_F\approx 0.1$ for $H=15\,\textrm{T}$. Estimates of the Landau parameter $F_0^\sigma$
are of the same order of magnitude\cite{Eshkol06,Zheng96} such that by tuning system parameters, 
the condition $u^{3/2}\ll |h|/E_F \ll 1$ seems to be within experimental reach.   

Interaction-induced magnetoresistance of a two-dimensional electron gas
in a transverse magnetic field has 
been calculated  by Gornyi and Mirlin~\cite{Gornyi03}.
They have
taken into account the same interaction processes as ZNA
but have considered also
the case of long-range  disorder, where their interaction
correction is exponentially suppressed by the disorder correlation function.
On the other hand, our interaction correction to the conductivity, which we have related
via the Langer-formula to the momentum-dependence of the self-energy in the clean limit, 
is not  exponentially suppressed for long-range disorder, indicating that also 
Gornyi and Mirlin did not take the non-analytic corrections
discussed in our work into account.

\section{Summary and conclusions}

In summary, we have shown that in a two-dimensional Fermi liquid
the quasi-particle residue and
the renormalized electronic Land\'{e} factor
exhibit a non-analytic magnetic field dependence proportional
to $| H  |$ at  zero temperature.
We have explicitly calculated the corresponding prefactors to second
order in the interaction.
To this order,  the terms linear in $| H |$ cancel 
in the renormalized  effective mass.
We have also shown that the magnetic field dependence
generated by the momentum-derivative of the electronic self-energy
gives rise to a non-analytic correction  to the density of states which 
is linear in $ | H|$ and
proportional to the square of the interaction in the weak coupling regime.

Using the Langer-formula~\cite{Langer61} for the
conductivity of a Fermi liquid,  we have also shown that
the momentum-derivative of the self-energy generates
a correction to the magnetoconductivity which is 
linear in $ | H |$  and
quadratic in the coupling constants  in both
the exchange and the Cooper channel.
We have compared our result with 
previous work by Zala, Narozhny and Aleiner \cite{Zala01} 
who have focused on the contribution from the exchange channel.
Whether the ZNA-formula implicitly takes into account the
non-analytic corrections due to the renormalization
of the current vertices
requires further investigations which are beyond the scope of this work.

While in this work we have focused on the non-analytic magnetic field dependence
of the self-energy at vanishing temperature,
in two-dimensional Fermi liquids
similar non-analyticities appear also as a function of temperature $T$.
In particular, we show in Appendix B that for $H=0$ and 
$T \ll E_F$ the factor $Y$ defined in  Eq.~(\ref{eq:selfexp}) has for a
constant bare interaction $U$ acting between electrons with
opposite spin the low-temperature expansion
 \begin{eqnarray}
 Y (T ) & = & 
  Y ( 0 )  + \tilde{Y}_1 T  + {\cal{O}} ( T^2 ), \; \; \; 
\tilde{Y}_1 = \frac{u^2}{4 E_F}.
 \label{eq:YT}
 \end{eqnarray}
A  linear temperature dependence of quasi-particle properties of two-dimensional
Fermi liquids has been discussed previously 
in Refs.~[\onlinecite{Galitski04,Chubukov08}] and is 
 closely related to terms of
order $ \omega T$ in the real part of the self-energy
on the mass shell discussed in Ref.~[\onlinecite{Chubukov03}].
Using again the Langer-formula (\ref{eq:Kubo}),  we find that the conductivity 
of  two-dimensional Fermi liquids in the ballistic regime  
exhibits a linear temperature dependence, in agreement with ZNA \cite{Zala01}.

Finally, let us point out two interesting extensions of our
calculations.
First of all, it would be interesting to calculate the 
coefficients $X_1$, $Y_1$, and $Z_1$ of the terms linear in 
the magnetic field in the low-energy expansion of the
electronic self-energy beyond the leading order
in the interaction. Such a calculation
should take vertex corrections and the interference
of the different low-energy scattering channels into account.
We believe that  the low-energy model with interaction (\ref{eq:Sphasespace})
is a good starting point for such a calculation.
It would also be interesting to revisit earlier calculations
of the magnetoconductivity in the presence of disorder~\cite{Zala01,Gornyi03},
consistently taking all sources of  non-analytic 
behavior by all low-energy scattering channels into account.
To determine the numerical value of the
coefficient $c_2$ in Eq.~(\ref{eq:condfullres}), a perturbative calculation
of the disorder averaged Kubo formula in the ballistic regime
which systematically includes all diagrams up to second order
in the interaction would be necessary.


\section*{ACKNOWLEDGMENTS}

We gratefully acknowledge the collaboration with Dmitrii Maslov 
during the early stages of this work and thank him for
numerous fruitful discussions and suggestions.
We also thank Boris Narozhny and Igor Gornyi for discussions, and
Tim  Herfurth and Andreas 
R\"{u}ckriegel for checking some of the numerical integrations.
Most of this work was carried out
during a sabbatical stay at the University of Florida, Gainesville; we would like to thank
the UF Physics Department for its hospitality.
PK and PL are grateful to the DFG for financial support via FOR 723.

\begin{appendix}

\renewcommand{\theequation}{A\arabic{equation}}

\section*{APPENDIX A: Derivation of the ZNA-formula to first
order in the interaction}
\setcounter{equation}{0}

\begin{figure}[tb]
  \centering
\vspace{7mm}
\includegraphics[width=80mm]{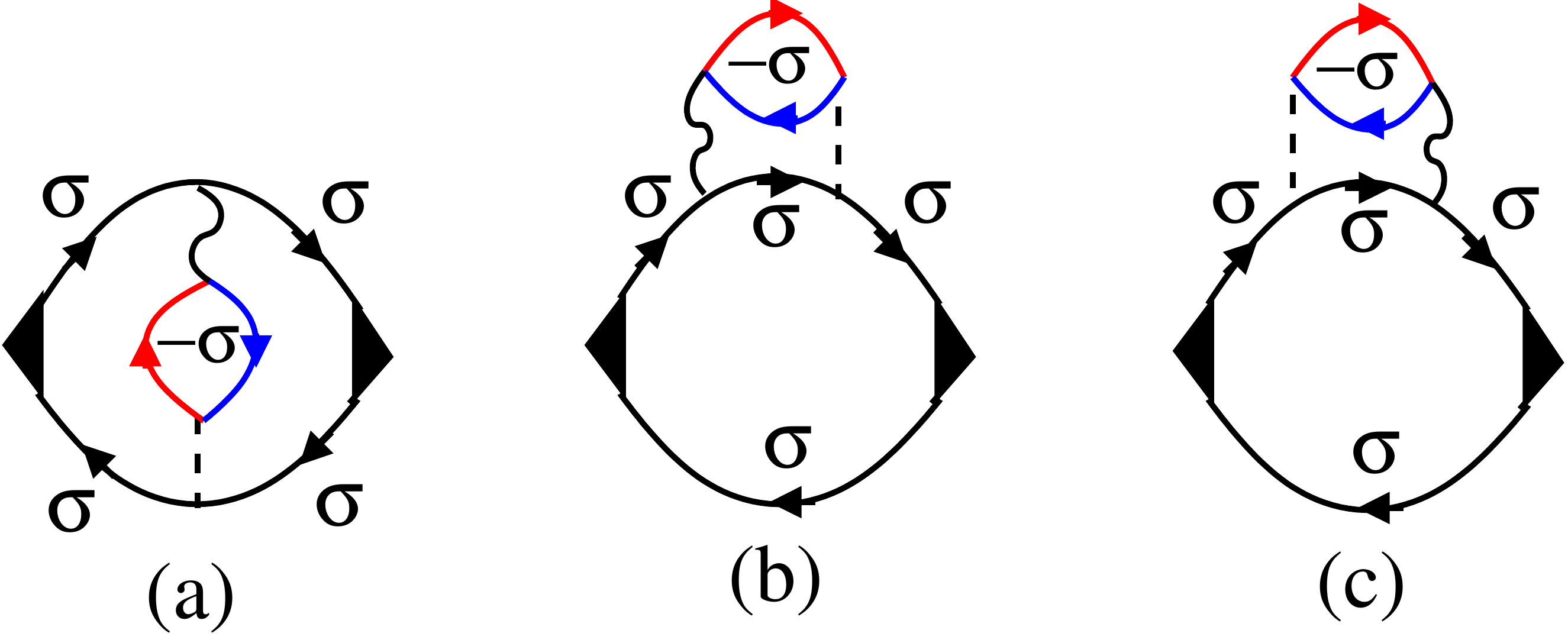}
\vspace{5mm}
  \vspace{-4mm}
  \caption{%
(Color online)
Corrections to  the disorder averaged Kubo formula for the conductivity
in the ballistic regime to first order in a momentum-independent interaction
which acts only between electrons with opposite spin-projection.
The black triangles are the bare current vertices. The other symbols are the same
as in Fig.~\ref{fig:impurity}.
The vertex correction (a) does not transfer any energy between
the two branches of the bubble 
and is not taken into account  in the Langer-formula (\ref{eq:Kubo}).
The diagrams (b) and (c) generate a non-analytic contribution to the renormalized
scattering time.
}
\label{fig:kubopert}
\end{figure}

In this appendix we re-derive the ZNA-formula  (\ref{eq:ZNA}) for the magnetoconductivity in the ballistic regime to leading (linear) order in the interaction.
For simplicity, we assume that the bare interaction $U$ 
is momentum independent and acts only between electrons with
antiparallel spin projection, see Eq.~(\ref{eq:action1}).
In the ballistic regime and to first order in the interaction
the magnetoconductivity is then determined
by the three diagrams  shown in Fig.~\ref{fig:kubopert}.
The diagrams (b) and (c) depend on the first order self-energy
diagrams shown in Fig.~\ref{fig:impurity} (b) and (c).
These diagrams give rise to non-analytic corrections to the
renormalized scattering time. To determine the corresponding contribution to the
conductivity, we can simply insert the relevant scattering time in the 
Drude formula. 
Assuming $\delta$-function correlated disorder, the 
sum of the two diagrams (b) and (c) in Fig.~\ref{fig:impurity}
yields the following first order contribution to the self-energy
of electrons with spin-projection $\sigma$,
 \begin{equation}
 \Sigma_1^{\sigma} ( k , i \omega )
 =  \frac{2 U }{2 \pi \nu \tau } 
 \int \frac{ d^2 q}{(2 \pi )^2 } \Pi^{ - \sigma, - \sigma}_0 ( \bd{q} , 0 )
 G_0^{\sigma} ( \bd{k} - \bd{q}  , i \omega ).
 \label{eq:sigma1}
  \end{equation}
The parallel-spin particle-hole bubble in a magnetic field is 
for vanishing frequency given by 
$\Pi^{  \sigma,  \sigma}_0 ( \bd{q} , 0 ) = - \nu P^{\sigma} ( q )$ with 
 \begin{equation}
  P^{\sigma} ( q )  =  1 - \frac{ 2 k_F^{\sigma}}{q}
 \Theta ( q - 2 k_F^{\sigma} ) \sqrt{ 
 \left( \frac{q}{2 k_F^{\sigma} } \right)^2 -1 },
 \label{eq:polstat}
 \end{equation}
where $k_F^{\sigma}  = k_F \sqrt{ 1 + \sigma h/E_F } \approx k_F [ 1 + \sigma h /(2 E_F) ]$.
After carrying out the angular integration in Eq.~(\ref{eq:sigma1}), we obtain
\begin{eqnarray}
 \Sigma_1^{\sigma} ( k , i \omega )
 & = &  \frac{2 u }{2 \pi \nu \tau } \int_0^{\infty} \frac{dq q}{2 \pi} P^{- \sigma} (q )
 \nonumber
 \\
 & \times &  \frac{ {\rm sgn} ( \xi_{{k}} + \frac{q^2}{2m} - \sigma h )}{
   \sqrt{ [ i \omega -  \xi_{{k}} - \frac{q^2}{2m} + \sigma h]^2 - (v_k q)^2 }}.
 \label{eq:sigma1b}
 \end{eqnarray} 
It turns out that the non-analytic magnetic-field dependence
of the self-energy (\ref{eq:sigma1b}) is due to the non-analyticity 
of the polarization bubble (\ref{eq:polstat}) at $q = 2 k_F^{\sigma}$.
It is therefore sufficient to replace Eq.~(\ref{eq:polstat}) 
by its approximation for $ | q - 2 k_F^{\sigma} | \ll 2 k_F^{\sigma}$,
 \begin{equation}
  P^{\sigma} ( q  )  \approx  1 -  \Theta ( q - 2 k_F^{\sigma} ) \sqrt{ \frac{ q - 2 k_F^{\sigma} }{
 k_F^{\sigma} } }.
 \end{equation}
To obtain the interaction correction to the scattering rate we need
 \begin{eqnarray}
 {\rm Im} \Sigma^{\sigma}_1 ( k_F^{\sigma} , i \eta ) & = & 
\frac{2 u }{2 \pi \nu \tau } \int_0^{\infty} \frac{dq q}{2 \pi} P^{- \sigma} (q )
 \nonumber
 \\
 & \times &
 {\rm Im} \frac{ {\rm sgn} ( \frac{q^2}{2m} - 2 \sigma h  ) }{ 
 \sqrt{ [ \frac{q^2}{2m} - 2 \sigma h - i \eta ]^2 - (v^{\sigma}_F q )^2 }}.
 \hspace{7mm}
 \end{eqnarray}
Setting $q = 2 k_F + p$ and approximating
 \begin{equation}
  \left[ \frac{q^2}{2m} - 2 \sigma h - i \eta \right]^2 - (v^{\sigma}_F q )^2 
 \approx  8 E_F [ v_F p - 2 \sigma h - i \eta ],
 \end{equation}
we obtain for  the spin-projection $\sigma = \uparrow$ to leading order in $h > 0$,
  \begin{eqnarray}
& & {\rm Im} \Sigma^{\uparrow}_1 ( k^{\uparrow}_F , i \eta ) 
 -  \left. {\rm Im} \Sigma^{\uparrow}_1 ( k_F^{\uparrow} , i \eta ) \right|_{ h=0}
 \nonumber
 \\
& = & - \frac{ 2 u}{ 2 \pi \nu \tau v_F}
  \int_{ - 2 h/v_F}^{ 2h/v_F} \frac{dp}{2 \pi} \sqrt{ \frac{v_F p+2 h }{2h - v_F p }}
=  - \frac{ u h }{ E_F \tau }. 
 \label{eq:sigmaup}
 \hspace{7mm}
 \end{eqnarray}
For the other spin projection we obtain for $h > 0$,
 \begin{equation}
 {\rm Im} \Sigma^{\downarrow}_1 ( k^{\downarrow}_F , i \eta ) 
 -  \left. {\rm Im} \Sigma^{\downarrow}_1 ( k_F^{\downarrow} , i \eta ) \right|_{ h=0} =0,
 \label{eq:sigmadown}
 \end{equation}
so that the renormalized  scattering rates are for $h >0$,
  \begin{equation}
 \frac{1}{ \tau_{\ast \uparrow}} = \frac{1}{\tau} - 2    {\rm Im} \Sigma^{\uparrow}_1 ( k^{\uparrow}_F , i \eta ) 
 = \frac{1}{\tau} \left[ 1 + \frac{2 u h }{ E_F }  \right],
 \end{equation}
and $1/ \tau_{\ast \downarrow} = 1/ \tau + {\cal{O}} ( h^2 )$.
The correction to the conductivity 
due to the diagrams shown in Fig.~\ref{fig:kubopert} (b) and (c) is therefore
 \begin{equation}
 \sigma (h )_{b+c} = \frac{ n e^{2}}{m} \frac{ \tau_{\ast \uparrow} + \tau_{\ast \downarrow}}{2}
 = \sigma_0 \left[ 1 -  u\frac{ | h |}{E_F} + {\cal{O}} ( u^2, h^2 ) \right].
 \label{eq:sigma3}
 \end{equation}
Keeping in mind that to leading order in the bare interaction 
we may identify $F_0^{\sigma} = - u $
and approximate $g ( F_0^{\sigma} )  \approx 1$ in Eq.~(\ref{eq:ZNA}),
the correction in Eq.~(\ref{eq:sigma3})
is a factor of two smaller than the corresponding correction
obtained from the weak-coupling expansion of the ZNA-formula (\ref{eq:ZNA}).
The missing contribution is due to the vertex correction diagram shown in
Fig.~\ref{fig:kubopert} (a), which yields the following
correction to the conductivity in the ballistic limit,
 \begin{eqnarray}
 \sigma ( h )_a & = & \frac{e^2}{ 2 \pi m^2}   \frac{ 2 U}{ 2 \pi \nu \tau }    \sum_{\sigma} 
 \int \frac{ d^2 k}{(2 \pi)^2 } \int \frac{ d^2 k^{\prime}}{(2 \pi )^2} \frac{ \bd{k} \cdot \bd{k}^{\prime}}{2}
 \nonumber
 \\
 & & \times  \Pi_0^{ - \sigma , - \sigma} (  \bd{k} - \bd{k}^{\prime} , 0 )
 G_0^{\sigma} ( \bd{k} , i \eta )  G_0^{\sigma} ( \bd{k} , - i \eta ) 
 \nonumber
 \\
& & \times
G_0^{\sigma} ( \bd{k}^{\prime} , i \eta )  G_0^{\sigma} ( \bd{k}^{\prime} , - i \eta ) .
 \label{eq:vertex}
 \end{eqnarray}
The correction linear in $ | h |$ can be extracted analytically and we obtain
 \begin{equation}
  \sigma ( h )_a - \sigma (0)_a = - \sigma_0 u \frac{| h |}{E_F}.
 \end{equation}
Combining this with Eq.~(\ref{eq:sigma3}) we finally obtain
 \begin{equation}
 \frac{\sigma ( h ) - \sigma (0)}{\sigma_0} = - 2 u \frac{ | h |}{E_F} + {\cal{O}} ( u^2 , h^2 ).
 \label{eq:condlin}
 \end{equation}
Keeping in mind that to leading order $F_0^{\sigma} = - u$ for our model, 
Eq.~(\ref{eq:condlin}) agrees
with the leading weak coupling behavior of the ZNA-formula (\ref{eq:ZNA}) in the
ballistic regime.

Finally, let us show that the vertex correction (\ref{eq:vertex})
is not taken into account in the  Langer-formula (\ref{eq:Kubo}).
As discussed in Sec.~\ref{sec:magnetoconductivity}, the
momentum-derivative of the self-energy in  the Langer-formula implicitly
takes vertex corrections due to interactions in the Kubo formula into account.
However, the vertex correction in Eq.~(\ref{eq:vertex}), which does not 
transfer any energy and can be viewed as an interference correction 
involving an effective  renormalized impurity line,
is not included in the Langer-formula.
To see this, note that
to first order in the interaction the velocity renormalization factor
in the Langer-formula can only be due to the momentum-derivative of the
self-energy $\Sigma_1^{\sigma} ( k , i \omega )$  
given in Eq.~(\ref{eq:sigma1}). However,   $\Sigma_1^{\sigma} ( k , i \omega )$   vanishes
for $1/ \tau \rightarrow 0$, so that in the ballistic limit the velocity renormalization
factor associated with the self-energy (\ref{eq:sigma1}) vanishes and therefore
does not contribute to the conductivity,
in contrast to Eq.~(\ref{eq:vertex}).

\renewcommand{\theequation}{B\arabic{equation}}

\section*{APPENDIX B: Non-analytic temperature dependence}
\setcounter{equation}{0}

In this appendix we briefly describe the derivation of 
the leading temperature dependence of the renormalization factor $Y$ 
for vanishing magnetic field given in Eq.~(\ref{eq:YT}). 
The derivation will be done along the lines of Ref.~[\onlinecite{Chubukov03}]. 
We begin by considering the exchange channel and express the 
antiparallel-spin particle-hole bubble through its 
spectral representation 
\begin{equation}
\Pi_0^{\uparrow\downarrow}(\bd{q},i\bar{\omega})=\frac{1}{\pi}\int_{-\infty}^\infty d\Omega\,\frac{{\rm Im}\Pi_0^{\uparrow\downarrow}(\bd{q},\Omega+i\eta)}{\Omega-i\bar{\omega}},
\label{eq:spectralpol}
\end{equation}
where $\eta>0$ is infinitesimal 
and the integral is now over real frequencies $\Omega$. 
The Matsubara sum in Eq.~(\ref{eq:selfb}) can then be performed. 
Using
${\rm Im}G_0(\bd{k},\omega+i\eta)=-\pi\delta(\omega-\xi_{\bd{k}})$ we 
find for the imaginary part of the self-energy
after analytic continuation to real frequencies,
 \begin{widetext}
 \begin{equation}
 {\rm Im}\Sigma_x( \bd{k} , 
 \omega + i \eta ) = \frac{U_x^2}{2}\int\frac{d^2q}{(2\pi)^2}\Theta(\Lambda_0-|\bd{q}|) 
 \int_{- \infty}^{\infty} d\Omega\, {\rm Im}\Pi_0^{\uparrow\downarrow}
 (\bd{q},\Omega + i \eta )\delta(\xi_{\bd{k}-\bd{q}}-\omega+\Omega)\left[\coth\left(\frac{\Omega}{2T}\right)-\tanh\left(\frac{\omega-\Omega}{2T}\right)\right].
\label{eq:selfspecim}
\end{equation}
The angular integration is now trivial due to the $\delta$-function while the remaining momentum integration can also be done. After performing a Kramers-Kronig transform
we obtain
for the singular contribution \cite{Chubukov03} to the real part of the self-energy,
\begin{equation}
{\rm Re}\Sigma_x( \bd{k} , \omega + i \eta ) = \frac{\nu U_x^2}{(2\pi)^3v_F^2}\mathcal{P}\int_{-\infty}^\infty\frac{d\omega^\prime}{\omega^\prime-\omega}\int_{-\infty}^\infty d\Omega\,\Omega\ln
 \left| \frac{v_F\Lambda_0}{2\Omega+\omega^\prime-\xi_{\bd{k}}}
 \right| \left[\coth\left(\frac{\Omega}{2T}\right)-\tanh\left(\frac{\omega^\prime+\Omega}{2T}\right)\right],
\label{eq:selfspecreal}
\end{equation}
\end{widetext}
where $\mathcal{P}$ denotes the principal part. Finally taking the derivative with respect to $\xi_{\bd{k}}$ and setting $ \omega= \xi_{\bd{k}} =0$ we can scale out the temperature dependence. By symmetrizing the integrand with respect to $\omega^\prime$, the remaining frequency integrals can then be done. 
We find that the exchange channel yields a contribution  $u_x^2/(8E_F)$
to the coefficient $\tilde{Y}_1$ in Eq.~(\ref{eq:YT}).

The contributions from the two other channels can be found in a similar way. 
In the Cooper channel the same steps lead to the following expression 
for the singular contribution to the imaginary part of the self-energy,  
 \begin{widetext}
\begin{equation}
{\rm Im}\Sigma_c( \bd{k} , \omega + i \eta ) = -\frac{\nu U_c^2}{(2\pi)^2}\int_0^{\Lambda_0} dq\,q \int_{-\infty}^\infty d\Omega\,\frac{\ln
 \left[ \frac{2\Omega_0}{|\Omega|+\sqrt{\Omega^2+(v_Fq)^2}} \right] }{\sqrt{(v_{\bd{k}}q)^2-(\xi_{\bd{k}}-\omega+\Omega)^2}}\left[\coth\left(\frac{\Omega}{2T}\right)-\tanh\left(\frac{\omega-\Omega}{2T}\right)\right].
\label{eq:selfspeccooper}
\end{equation}
Differentiating with respect to $\xi_{\bd{k}}$, setting then
 $\omega = \xi_{\bd{k}} =0$, and using the Kramers-Kronig transform we find
for the real part, 
\begin{eqnarray}
\left. \frac{\partial{\rm Re}\Sigma_c(\bd{k}, i \eta)}{\partial\xi_{\bd{k}}}\right|_{\xi_{\bd{k}}=0} &=& \frac{\nu U_c^2}{(2\pi)^3v_F^2}\int_0^{v_F^2\Lambda_0^2} dx\,\int_{-\infty}^{\infty}d\Omega\,\ln \left[ \frac{2\Omega_0}{|\Omega|+\sqrt{\Omega^2+x}} \right] \nonumber \\
& & \times \mathcal{P}\int_{-\infty}^\infty\frac{d\omega^\prime}{\omega^\prime}\frac{\omega^\prime-\Omega}{\left[x-(\omega^\prime-\Omega)^2\right]^{3/2}}\left[\coth\left(\frac{\Omega}{2T}\right)-\tanh\left(\frac{\omega^\prime-\Omega}{2T}\right)\right], 
\label{eq:Ycooper}
\end{eqnarray}
\end{widetext}
with $x=v_F^2q^2$. Since the integrand is odd under the simultaneous shift of $\Omega\rightarrow -\Omega$ and $\omega^\prime\rightarrow -\omega^\prime$ the integral vanishes. Finally for vanishing magnetic field the contribution from the forward scattering channel 
to $Y$  can be taken into account by simply replacing $u_x \rightarrow u_f$
in the result from the exchange scattering channel, so that for $u_x = u_f$
forward and exchange scattering channels yield identical contributions to $Y$.
Collecting all terms and setting $u_x=u_f=u_c=u$ we finally 
arrive at Eq.~(\ref{eq:YT}).

\end{appendix}

\end{document}